\documentclass{elsart4}

\usepackage{amssymb}

\usepackage{graphics}
\usepackage{graphicx}
\usepackage{epsfig}
\usepackage{amssymb}
\journal{Physica B}

\begin{document}

\begin{frontmatter}

\title{Theoretical study of the magnetism in the incommensurate phase of TiOCl}
\author{\corauthref{cor1} D.M. Mastrogiuseppe},
\author{M.E. Torio},
\author{C.J. Gazza},
\author{A.O. Dobry} \address{Facultad de Ciencias Exactas Ingenier\'ia y Agrimensura, \\
Universidad Nacional de Rosario and Instituto de F\'isica Rosario,\\
Bv. 27 de Febrero 210 bis., 2000 Rosario, Argentina}

\corauth[cor1]{Corresponding author. Tel.: +54-341-4853200 ext. 127; fax: +54-341-480-8584.
\\E-mail address: mastrogiuseppe@ifir-conicet.gov.ar}

\begin{abstract}
Going beyond a recently proposed microscopic model \cite{diegoyariel}
for the incommensurate transition in the spin-Peierls TiOX (X=Cl, Br) compounds,
in the present work we start by studying the
thermodynamics of the model with XY spins and adiabatic phonons. We find that 
the system enters in an incommensurate phase by a first order transition at 
a low temperature $T_{c1}$. At a higher temperature $T_{c2}$ a
continuous transition to a uniform phase is found. Furthermore, we
study the magnetism in the incommensurate phase by Density Matrix
Renormalization Group (DMRG) calculations on a 1D Heisenberg model
where the exchange is modulated by the incommensurate atomic position pattern.
When the wave vector $q$ of the modulation is near $\pi$, we find local magnetized zones
(LMZ) in which spins abandon their singlets as a result of the domain walls induced by
the modulated distortion. When $q$ moves far away enough from $\pi$, the LMZ disappear and the system
develops incommensurate magnetic correlations induced by the
structure. We discuss the relevance of this result regarding previous and future experiments in TiOCl.
\end{abstract}

\end{frontmatter}

\section{Introduction}
The interest on spin-Peierls systems \cite{reviewSP} has been renewed with 
the recent discovery of a dimerization in the TiOX (X = Cl, Br) compounds family.
In them, the magnetically active Ti ions connected by O ones are arranged in a bilayer
structure.
Ti ion positions on a layer are shifted with respect to the other
neighboring layer, forming something like an anisotropic triangular structure.
Initially postulated for a \textit{resonating valence bond} state \cite{beynon},
TiOCl turned out to be mainly a one-dimensional magnetic system \cite{shaz}
with the Ti d$_{xy}$ orbitals pointing towards each other in the crystallographic
$b$ direction \cite{seidel}.
The high temperature magnetic susceptibility is well described by the
Bonner-Fisher curve, indicating a nearest neighbor magnetic exchange
$J\sim 660K$ \cite{seidel}.
The peculiar ingredient in the TiOCl phase diagram is the appearance of an
incommensurate intermediate phase between the high temperature uniform phase
and the low T dimerized phase \cite{seidel,imai} commonly found in a regular spin-Peierls system.
The transition temperatures were experimentally found to be $T_{c1}\sim 66K$ and
$T_{c2}\sim 92K$. In Ref. \cite{imai} a very large energy
gap of about $430K$ in the low temperature phase and a pseudo spin gap below
$135K$ have been also reported.

The origin of the intermediate phase remains controversial and not
well understood yet. The shifted positions of a Ti ion in a chain between
the two neighboring chains, make it plausible to
entail that some type of competition between the
in-chain and out-of-chain interactions could be the origin of the
incommensurate phase.

Some of us recently proposed a microscopic mechanism for
the incommensurate phase \cite{diegoyariel}, where the  antiferromagnetic
Ti chains are immersed in the phonon bath of the bilayer structure.
By using the Cross-Fisher theory \cite{CF}, it was shown that the
geometrically frustrated character of the lattice is 
responsible for the structural instability, leading the chains
to an incommensurate phase without an applied magnetic field.
For TiOCl, our results showed consistency with the
temperature dependence of the phonon frequencies and the value of
the incommensuration vector at the transition temperature.
Moreover, we found that the dynamical structure factor shows a
progressive softening of an incommensurate phonon near the zone
boundary as the temperature decreases, along with a broadening of
the peak.
These features are in agreement with the
experimental inelastic X-ray measurements \cite{abel}.

The purpose of this work is double. As a first step we present the phase diagram as a function of the temperature
of the model proposed in Ref. \cite{diegoyariel}. 
On the other hand, we theoretically analyze the evolution of the magnetism inside the incommensurate phase
by solving a one-dimensional Heisenberg model with a distortion-dependent exchange, using an incommensurate wave as 
the modulation of the Ti positions. This study is performed using Density Matrix Renormalization Group (DMRG) calculations.

\section{The model and its phase diagram}

The model proposed in Ref. \cite{diegoyariel} contains the relevant
magnetic interaction, the phonons and the spin-phonon coupling. It
includes the Ti atoms on the bilayer structure which interact by harmonic
forces. 
For simplicity, we consider ionic movements only in the direction of the magnetic chains.
As the measured magnetic susceptibility of TiOCl is well reproduced by a 1D
Heisenberg model, the assumption that the Ti atoms are magnetically
coupled only along the $b$ direction was taken into account. Moreover, as a direct
exchange seems to be the dominant Ti-Ti interaction, it is reasonable to consider that
$J$ will be modulated by the movement of nearest neighbor Ti ions in the chain direction.
Our spin-phonon Hamiltonian reads:

\begin{eqnarray}
\label{hspinph}
H&=&H_{ph}+H_{sph},\\
H_{ph}&=&\sum_{i,j}{\frac{P_{i,j}^2}{2m}} +  \sum_{i,j}\lbrace\frac{K_{in}}{2}(u_{i,j}-u_{i+1,j})^2\nonumber\\
&+&\frac{K_{inter}}{2}\left[( u_{i,j}-u_{i,j+1})^2 +( u_{i,j}-u_{i+1,j-1})^2\right]\rbrace,\nonumber\\
H_{sph}&=&\sum_{i,j} J_{i,j} \textbf{S}_{i,j} \cdot
\textbf{S}_{i+1,j},
\label{hspin}
\end{eqnarray}
\noindent where $u_{i,j}$ is the displacement of the atom $i$ of the chain $j$
along the direction $b$ of the magnetic chains and the $P_{i,j}$ are their 
respective momenta. $\textbf{S}_{i,j}$ are spin-$\frac12$ operators with exchange
constant $J_{i,j}=J[1 + \alpha (u_{i,j}-u_{i+1,j})]$ along the b-axis of a non-deformed underlying
lattice. $K_{in}$ and $K_{inter}$ are the harmonic force constants.

The Hamiltonian (\ref{hspinph}) contains variables of different dimensionality: whereas the spins are one-dimensional, phonons are two-dimensional. 
An exact solution to this model at finite temperature is a formidable task, therefore we make two approximations to obtain the phase diagram. 
We consider only the XY coupling in the spin-spin interaction and phonons are treated in the adiabatic approximation. Spin variables are converted 
to spinless fermions by means of the Jordan-Wigner transformation. The resulting pseudofermion-phonon model is solved in a finite system by 
diagonalizing the one-electron problem in the background of a deformed lattice (let us call each eigenvalue
$\lambda^\gamma, \gamma=1,...,N_{s}$, where $N_{s}$ is the the number of sites in each chain).

Now we consider deformations of the form:
\begin{eqnarray}
u_{i,j}&=&u_0 \cos({\bf q \cdot R}_{i,j}-\pi/4),
\label{uij}
\end{eqnarray}
where the wave vector ${\bf q}=(0,q)$ is constrained to the chain direction and ${\bf R}_{i,j}$ is the position vector in the two-dimensional Bravais lattice. We have chosen the $\pi/4$ phase to have a continuous energy at $q=\pi$.
The total free energy per site is:
\begin{eqnarray}
f=\frac{u_0^2 M \omega^2({\bf q})}{2} - \frac{T}{N_s} \sum_{\gamma=1}^{N_{s}}\ln(1+e^{-\lambda^\gamma/T}),
\end{eqnarray}
where $T$ is the temperature in units of $J$
and $\omega^2({\bf q})=\frac{4}{M} \left[K_{in}\sin^2(\frac{q_y}{2})+K_{inter} (1-\cos(\frac{q_x}{2})\cos(\frac{q_y}{2}))\right]$ 
is the dispersion relation of the two-dimensional phonons.

\begin{figure}
\centering
\includegraphics[width=7cm]{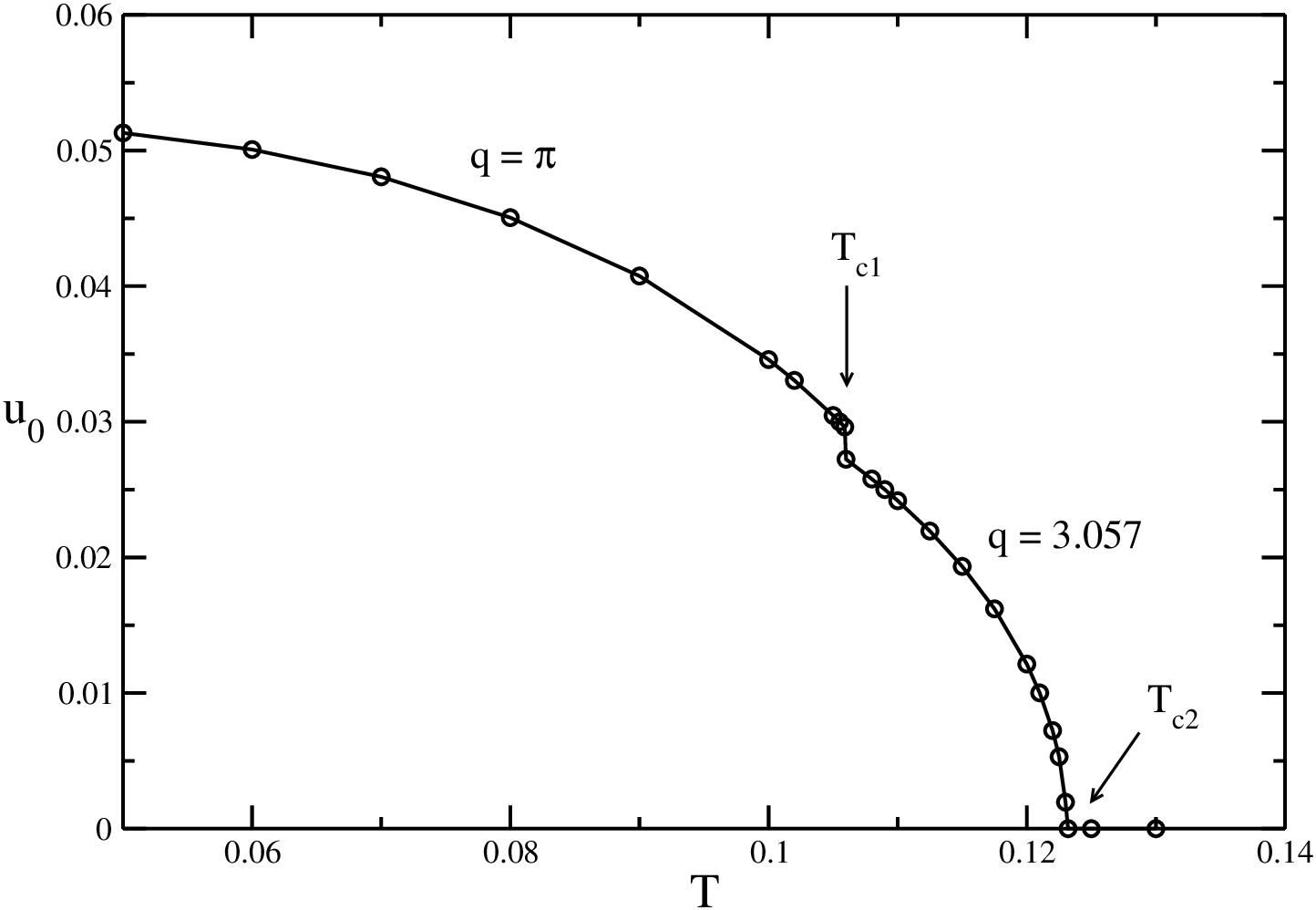}
\caption{Phase diagram in the plane $u_0, T$ of the model \ref{hspinph} as discussed in the text.
Parameters are $\alpha=3$ and $\frac{K_{inter}}{K_{in}}=1.8$. We identify each phase with the value of the
wave vector of the pattern that minimize the free energy.}
\label{phdiag}
\end{figure}

We have minimized $f$ with respect to $u_0$ and $q$ at different temperatures. The results are shown in Fig. \ref{phdiag}. At low temperatures a
dimerized phase is found. This is the usual result for an isolated spin-Peierls chain. At the first transition temperature $T_{c1}$, $u_0$ jumps 
discontinuously and $q$ moves away from $\pi$. This is a structural incommensurate phase which arises as a competition between the tendency to the 
dimerization of each chain and the frustration in the elastic interchain coupling.
The value of $q$ remains constant up to a second temperature $T_{c2}$ where $u_0$ vanishes and the system becomes uniform.
The sequence of the two transitions and the first order character of the first transition \cite{abel,krimmel} experimentally found
in TiOCl can therefore be accounted by our model. However the precise fitting of the parameters is a quite complex task which would 
need the inclusion of the z-component of the spin interaction and the non adiabatic effects of the phonons \cite{DCR}.

\section{Magnetic structure in the incommensurate phase}

The feature that makes the compound under study different from an
usual spin-Peierls system is the appearance of an incommensurate phase
without an applied magnetic field.
It is therefore quite important to characterize the magnetism in
this phase. As no magnetic interchain interaction was
included in the model (\ref{hspinph}),
we will consider a one-dimensional Heisenberg model as given by
(\ref{hspin}) with deformation fixed by (\ref{uij}). The model
will be solved by DMRG. We will focus on the local magnetization
and spin-spin correlation functions.

\begin{figure}
\centering
\includegraphics[width=7cm]{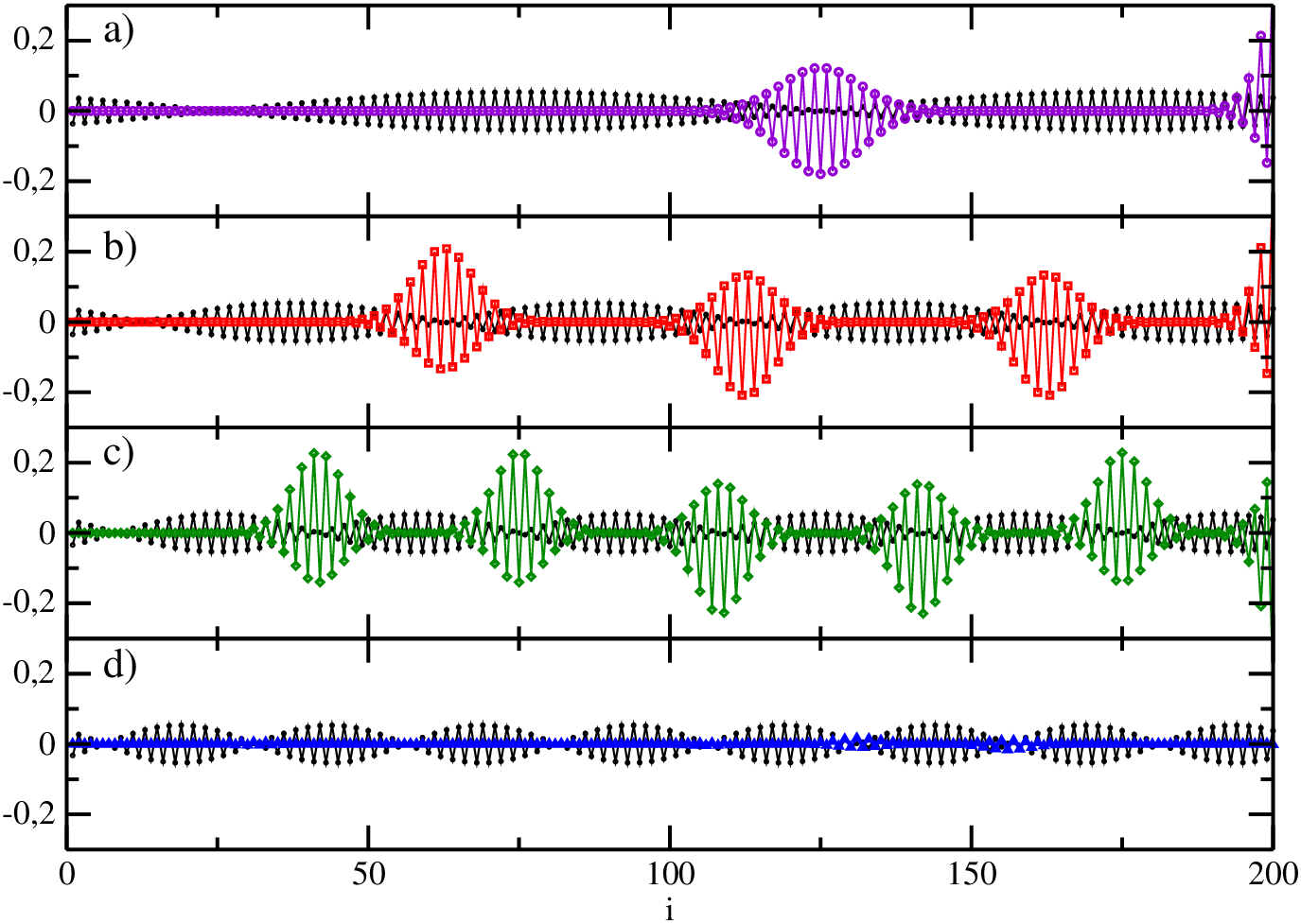}
\caption{DMRG results for a lattice of $N_s=200$. We
show in black the atomic positions for different $q_l=\pi(1-
\frac{2 l}{N_s})$, with $l=1,2,3,4$ from top to bottom, and $u_0 =0.053$.
Superimposed, we plot the $<S_i^z>$ to show the appearance of the LMZ.}
\label{correla}
\end{figure}

As a reference state to discuss the results, we take the
dimerized phase because the structural deformation (\ref{uij})
with $q$ near $\pi$ can be thought as a smooth modulation of the
dimer pattern. In a dimerized phase the spins are paired in
singlets and the local magnetization is zero. There is a gap in the
spectrum and the magnetic correlation declines exponentially. When the
pattern (\ref{uij}) is included with $q_l=\pi(1-\frac{2 l}{N_s})$ to have an 
integer number of wavelengths in the chain, the 
dimerization phase changes its sign $2 l$ times as we
travel along the chain from one extreme to the other (see the gray line
of Fig. \ref{correla}). When $l$ is small
enough, the situation has some similarities with the domain walls
(or solitons) appearing in the presence of a high magnetic field in regular
spin-Peierls systems \cite{Feiguin}.
We expect that for each zero of (\ref{uij}) free LMZ should appear. 
This is in fact the situation we found
in Fig. \ref{correla}a, b and c  where the mean value $<S_i^z>$ is
shown. We see the appearance of LMZ in all the zeros of the
deformation except in the first, which is greatly affected by the
proximity to the edge. These LMZ  are seen as clouds of polarized
spins emerging from the singlet background. Moreover, they are
uncorrelated among them because their distances are longer than the
correlation length of the dimerized state. They are randomly
orientated. We expect that the magnetic susceptibility will be
affected by these LMZ. It should increase compared to the
dimerized phase. This feature has been experimentally observed in
TiOCl \cite{abel}.

When $q_l$ moves further away from $\pi$, the zeros of (\ref{uij})
fall within a distance of the order of the correlation length and the LMZ disappear
as seen in Fig. \ref{correla}d. The effect of the structural
incommensuration on the magnetism is more subtle and a careful
analysis of the magnetic correlation function is necessary.

\begin{figure}
\centering
\includegraphics[width=7cm]{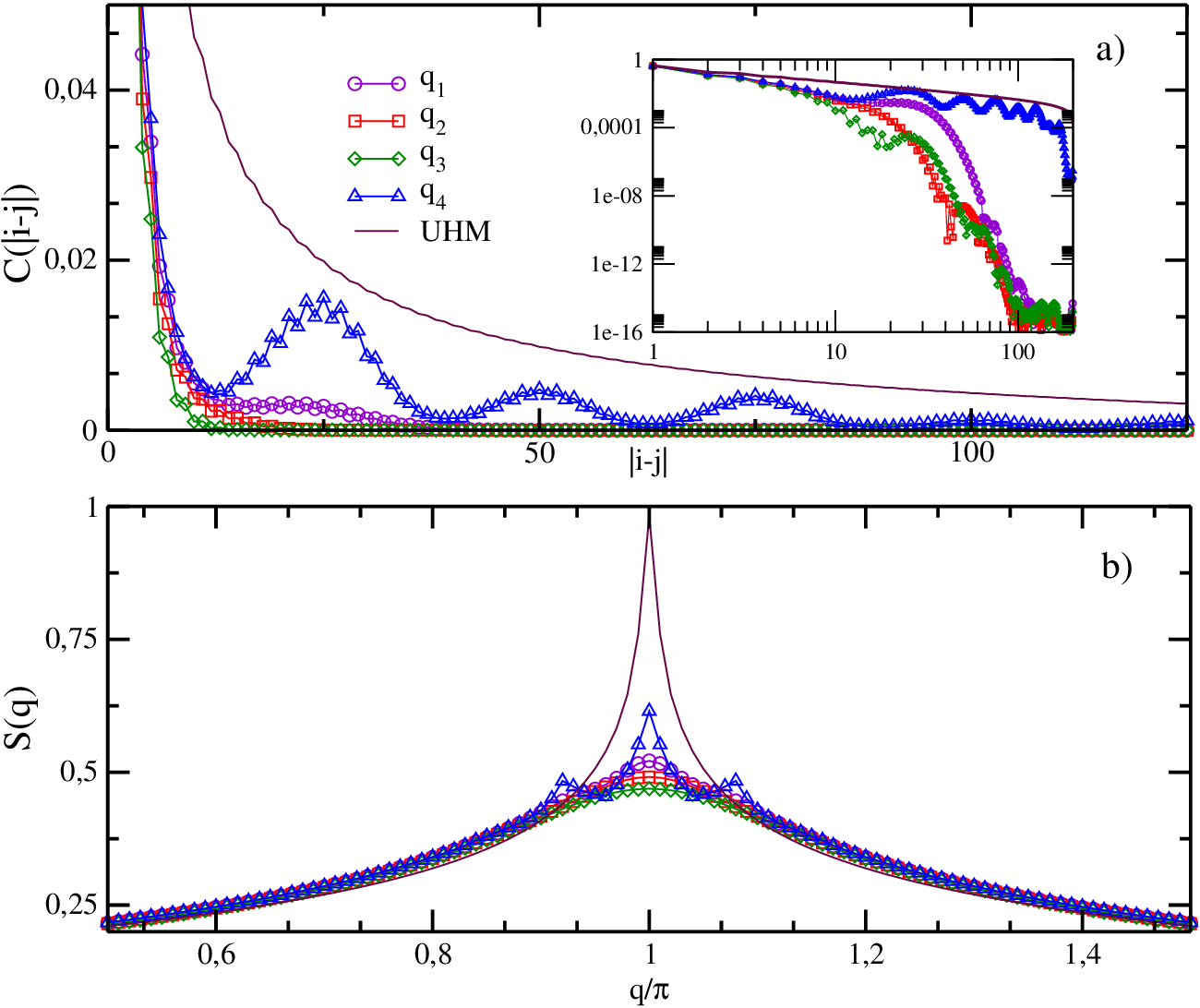}
\caption{(a) DMRG results for the spin-spin correlation functions
$C(\mid i-j \mid)$ for deformations with the $q_l$ as given
in Fig. \ref{correla}. We also show the correlation of a UHM 
for comparison. In the inset we show a log-log plot of $C(\mid  i-j \mid)$.
(b) Fourier transforms of the correlation functions of (a).}
\label{correlb}
\end{figure}

In Fig. \ref{correlb} we show the spin-spin correlation
function $C(\mid i-j \mid) \equiv \mid<{\bf S}_i\cdot{\bf S}_j>-<{\bf
S}_i><{\bf S}_j>\mid$ and its Fourier transform (the
static magnetic structure factor $S(q)$). For the three $q$ where the LMZ
appear, the correlation decreases exponentially to zero. Moreover, it falls a bit faster 
the farther $q$ is from $\pi$.
This seems to be a finite size effect because the modulation interferes with
the decrease of the spin correlation of the dimerized zones when a
zero appears. So we expect that in the thermodynamic
limit the dimerized region will be large enough for the
correlation to decrease as in the dimerized phase.

A more dramatic change in the behavior of \mbox{$C(\mid i-j \mid)$} is
found for $q_4$ where the LMZ do not appear. In this case the
correlation decreases more slowly than in the previous cases.
For comparison we show the result of a uniform Heisenberg model (UHM) where
the quasi-long range antiferromagnetic order is expected.
Note that $C(\mid i-j \mid)$ for $q_4$ approaches the one of the UHM, a quite different behavior than 
that of $q_1$, $q_2$ and $q_3$. We will undertake a careful analysis
of this long distance behavior for different modulations in a forthcoming paper.

More importantly, there is a modulation of the long distance behavior
with half the wavelength of that which has the imposed structure.
The magnetic structure can no longer be interpreted in terms of
a weak perturbation of the dimerized state. Instead, the magnetism
becomes incommensurate as a response to the modulated exchange. In
some sense, the situation is the dual to the one recently studied in
some multiferroic materials \cite{multiferro}. There, an helical 
magnetic structure induces a lattice distortion which finally
produces the ferroelectricity, whereas in our case the
lattice distortion is the driving force which produces an
incommensuration of the magnetic state.

These features are even more apparent in the structure factor $S(q)$
shown in Fig. \ref{correlb}b. For $q_1$, $q_2$ and $q_3$ there is 
a single maximum at $q_{max}=\pi$ whose 
intensity decreases as $q$ moves away from $\pi$. This maximum is rounded 
as in a pure dimerized chain \cite{Chitra}. It can be fitted near $q=\pi$
by the law $\frac{1}{(q-\pi)^2+\xi^{-2}}$ (corresponding to a free massive 
boson propagator \cite{whiteaffleck})
with $\xi \sim 3.3$. This asserts our previous statement that
imposing distortions with wavector $q_l$ near enough to $\pi$, the long 
range magnetic correlation behaves as in a dimerized chain, decreasing exponentially
with a correlation lenght of the order of $\xi$. For $q_l=q_4$ the behavior changes qualitatively. 
Two additional maxima appear at one side and 
the other of the highest peak at $\pi$, corresponding to the long distance magnetic modulation found in 
$C(\mid i-j \mid)$. Moreover, the peak at the center is not longer rounded but it behaves similarly to the UHM.
This signals a tendency towards a quasi-long range incommensurate magnetic correlation of the system. 
Let us make a brief discussion about the application of these results to TiOCl.
As measured from X-ray experiments \cite{abel,krimmel}, the modulation wave
vector is $q\sim 3.05$ at $T_{c2}$ and it moves slightly to $\pi$ when decreasing the
temperature towards $T_{c1}$ where it jumps discontinuously to $\pi$. So we expect
a change of the magnetic behavior inside the incommensurate phase from a structure with
LMZ to another without LMZ but with an incommensurate correlation function. This should be
checked in future magnetic experiments.
\newline

In summary we have found that a system with one-dimensional
antiferromagnetic chains coupled to two-dimensional phonons in an
anisotropic triangular lattice accounts for the appearance of the
incommensurate phase in TiOCl. The magnetism in this intermediate
state includes uncorrelated LMZ near the dimerized phase, and incommensurate 
correlations with no LMZ towards the uniform phase. In the last case, the
modulation found in the correlation function is related to the modulation
of the imposed distortion by a factor of two in their wavelengths. 
\newline

This work was supported by ANPCyT (PICT 1647), Argentina.

\end{document}